\documentclass[12pt]{article}
\usepackage[margin=1in]{geometry}
\usepackage{authblk}

\usepackage{graphicx}
\usepackage{booktabs}
\usepackage[hidelinks]{hyperref}

\usepackage{cite}

\usepackage{url}
\usepackage{amsmath}

\begin{document}

%

%
%
%

\title{Steering through Time: Blending Longitudinal Data with Simulation to Rethink Human-Autonomous Vehicle Interaction}

\author[1]{Yasaman Hakiminejad}
\author[1]{Shiva Azimi}
\author[2]{Luis Gomero}
\author[3]{Elizabeth Pantesco}
\author[3]{Irene P. Kan}
\author[2]{Meltem Izzetoglu}
\author[1]{Arash Tavakoli}

\affil[1]{Department of Civil and Environmental Engineering, Villanova University, Villanova, PA, USA\\
\texttt{\{yhakimin, sazimi, arash.tavakoli\}@villanova.edu}}

\affil[2]{Department of Electrical and Computer Engineering, Villanova University, Villanova, PA, USA\\
\texttt{\{lgomero, meltem.izzetoglu\}@villanova.edu}}

\affil[3]{Department of Psychological and Brain Sciences, Villanova University, Villanova, PA, USA\\
\texttt{\{elizabeth.pantesco, irene.kan\}@villanova.edu}}

\date{}

\maketitle
\begingroup
\renewcommand\thefootnote{}
\footnotetext[0]{Y. Hakiminejad, S. Azimi, and L. Gomero contributed equally to this manuscript.}
\endgroup

\begin{abstract}
As semi-automated vehicles (SAVs) become more common, ensuring effective human-vehicle interaction during control handovers remains a critical safety challenge. Existing studies often rely on single-session simulator experiments or naturalistic driving datasets, which often lack temporal context on drivers’ cognitive and physiological states before takeover events. This study introduces a hybrid framework combining longitudinal mobile sensing with high-fidelity driving simulation to examine driver readiness in semi-automated contexts. In a pilot study with N=38 participants, we collected 7 days of wearable physiological data and daily surveys on stress, arousal, valence, and sleep quality, followed by an in-lab simulation with scripted takeover events under varying secondary task conditions. Multimodal sensing—including eye tracking, fNIRS, and physiological measures—captured real-time responses. Preliminary analysis shows the framework’s feasibility and individual variability in baseline and in-task measures; for example, fixation duration and takeover control time differed by task type, and RMSSD showed high inter-individual stability. This proof-of-concept supports the development of personalized, context-aware driver monitoring by linking temporally layered data with real-time performance.

\end{abstract}



%

\section{Introduction}
Distracted and drowsy driving contribute to nearly 8\% of fatal crashes in the U.S. \cite{NHTSA2024DistractedDriving}. To mitigate risks from human error, semi-automated vehicles (SAVs) are increasingly deployed \cite{ding2020avt}. These systems share control with drivers and can issue takeover requests at any time, requiring drivers to respond in a timely fashion \cite{morales2020automated}. Delayed responses have led to fatal crashes, prompting regulations for attention monitoring, such as detecting hand placement on the steering wheel \cite{fu2024advancements}. Driver Monitoring Systems (DMS) assess cognitive and behavioral states—such as attention, fatigue, and distraction—and deliver timely interventions \cite{kumar2025comprehensive}. By detecting when drivers are ``out of the loop,'' DMS help counteract complacency and overtrust. 


DMS research typically uses naturalistic driving studies or controlled simulator experiments. Naturalistic studies capture authentic driver behavior over extended periods in real-world contexts \cite{eenink2014udrive}, while simulators offer precise control over variables such as takeover timing and cognitive load—ideal for isolating behavioral and physiological responses. However, naturalistic studies are resource-intensive with limited control \cite{ziakopoulos2020critical}, whereas simulator studies often rely on single-session data and overlook variability in drivers’ baseline psychophysiological states. This limits our understanding of how fluctuating factors like stress, sleep, or cognitive fatigue shape takeover readiness. For example, Magaña et al. \cite{magana2020effects} found that elevated pre-experiment stress impaired reaction time, lane-keeping, and hazard response in simulated driving. These findings highlight a gap: current DMS may underrepresent upstream psychophysiological influences, limiting real-time adaptability.

Advances in mobile sensing and wearable technologies enable combining real-world ecological validity with experimental control \cite{hong2024advances}. Lightweight, affordable smartwatches and ubiquitous smartphones enable continuous, unobtrusive collection of physiological and behavioral data at scale \cite{pinge2024detection}, capturing natural fluctuations in stress, sleep, activity, and emotional arousal over time. These capabilities support individualized baseline models and open opportunities for adaptive systems in transportation, health, and human–machine interaction.

This paper introduces a hybrid framework that combines high-fidelity driving simulation with mobile sensing and daily experience sampling. By capturing both baseline psychophysiological states and real-time responses during takeover scenarios, the approach enables a more ecologically valid view of driver behavior in semi-automated vehicles. We report results from a pilot study (N = 38) to demonstrate feasibility and highlight individual variability in cognitive, physiological, and behavioral responses. 


\subsection{Objectives and Hypotheses}

The objective of this study is to develop and evaluate a hybrid framework that integrates longitudinal mobile sensing, daily experience sampling, and high-fidelity simulation to study driver behavior in semi-automated vehicles. By capturing pre-experiment psychophysiological states and linking them to in-task responses—such as takeover control, gaze patterns, and cortical activation—this framework aims to characterize individual variability and support future context-aware driver monitoring systems.

Subsequently, we explore three key hypotheses: (H1) Drivers show stable individual differences in psychobehavioral states during the pre-experiment period, as captured by mobile sensing and experience sampling; (H2) Engagement in cognitively demanding secondary tasks (e.g., phone conversation, n-back) impairs takeover performance compared to no-task conditions; and (H3) The nature of the takeover context (e.g., crash vs. unexpected pedestrian) shapes drivers’ physiological, neural, and behavioral responses.









\section{Methodology}

\subsection{Participant Recruitment and Demographics} \label{lab:particiant}

This study was approved by Villanova University’s Institutional Review Board (IRB-FY2024-59). We recruited 38 participants via university mailing lists, flyers, and word of mouth. Eligible participants were 18 or older with a valid driver’s license. They could choose to enroll in the full study (longitudinal monitoring + simulation) or the driving-only portion. Demographic details are shown in Table~\ref{tab:demo}.

\begin{table}[ht]
\centering
\caption{Participant Demographics}
\label{tab:demo}
\resizebox{0.5\textwidth}{!}{%
\begin{tabular}{ll}
\toprule
\textbf{Characteristic} & \textbf{Summary} \\
\midrule
\textbf{Age (years)} & $27.8 \pm 8.0$ \\
\midrule
\textbf{Gender} & \\
\hspace{1em}Male & 21 \\
\hspace{1em}Female & 16 \\
\hspace{1em}Other / Unspecified & 1 \\
\midrule
\textbf{Ethnicity} & \\
\hspace{1em}White & 25 \\
\hspace{1em}Asian & 5 \\
\hspace{1em}Black or African American & 3 \\
\hspace{1em}White, Other & 1 \\
\hspace{1em}White, Asian & 1 \\
\hspace{1em}Prefer not to say & 1 \\
\hspace{1em}Other / Unspecified & 2 \\
\bottomrule
\end{tabular}%
}
\end{table}

\subsection{Study Overview and Design} \label{sec:study_design}

This study employed a $2 \times 3$ within-subjects design to examine driver responses under varying cognitive and contextual demands. Each participant experienced six conditions combining two \textbf{takeover contexts}—an unexpected pedestrian scenario (dynamic) and a crash scene (static)—with three levels of \textbf{secondary task engagement}: no task (control), a 2-back working memory task, and a naturalistic phone conversation. This design enabled analysis of intra-individual variability in physiological, neural, and behavioral responses to takeover requests.

The study consisted of two phases: (1) a 7-day longitudinal pre-experiment phase involving wearable sensing and daily self-reports, and (2) an in-lab simulation session where participants experienced takeover events while fitted with multiple sensing systems.

\subsection{Participants and Procedure}

Participants were recruited from the university community, provided informed consent, assigned a unique ID, and chose either the full study (longitudinal + simulator) or simulator-only portion.

In Visit 1, full-study participants were fitted with an Empatica EmbracePlus wristband and installed two mobile apps: the Empatica app for passive data syncing and the Ethica app for experience sampling. They were instructed to wear the device consistently (excluding water exposure) and respond to surveys upon notification.

Seven days later, all participants returned for Visit 2 in the Driving Simulator Room at Villanova University. Upon arrival, they were fitted with additional sensing devices and completed three experimental driving sessions. Each session included an autonomous driving phase followed by a takeover request (pedestrian or crash event). Secondary task condition was manipulated within subjects, such that all participants experienced all three secondary task conditions (no task, 2-back task, and phone conversation) across different driving sessions. The order of secondary task conditions was randomized to mitigate order effects. The two takeover contexts were presented in a fixed order for all participants, with the unexpected pedestrian scenario occurring first, followed by the crash scenario, and were not counterbalanced. After each event, participants pulled over and completed the NASA-TLX workload assessment \cite{hart1988development}.

At the end of Visit 2, participants completed a post-experiment questionnaire that included the Big Five Inventory (Extra-Short Form) \cite{soto2017short}, a sleep quality index \cite{krystal2008measuring}, and a driving behavior inventory adapted from validated instruments \cite{bygren1974driver}. Demographic information was also collected. The full session lasted approximately one hour.

\subsection{Longitudinal Monitoring}

\subsubsection{Wearable Physiological Sensing}

During the 7-day pre-experiment period, participants wore the Empatica EmbracePlus wristband, which continuously recorded four physiological signals: electrodermal activity (EDA, 4 Hz) for sympathetic arousal; photoplethysmography (PPG, 64 Hz) for heart rate and interbeat intervals; a 3-axis accelerometer (32 Hz) for activity levels; and an infrared thermopile (4 Hz) for peripheral skin temperature. Data were synced via the mobile app and stored on a secure cloud using anonymized IDs.

\subsubsection{Experience Sampling Method (ESM)}

\begin{figure}
    \centering
    \includegraphics[width=1\textwidth]{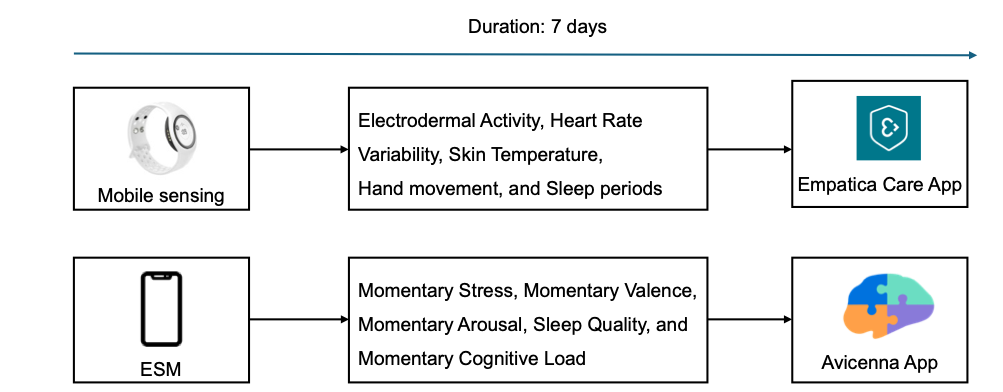}
    \caption{Longitudinal data collection framework}
    \label{fig:long}
\end{figure}

Participants received four self-report surveys per day via the Ethica app during the 7-day pre-experiment period, resulting in up to 28 self-report samples per participant. As expected in experience-sampling studies, some responses were missing due to noncompliance. Surveys measured stress (10-point scale) \cite{karvounides2016three}, emotional valence and arousal (7-point scales) \cite{mauss2009measures}, perceived mental workload \cite{hart1988development}, and sleep quality (5-point scale in morning surveys) \cite{krystal2008measuring}. These self-reports provided ecologically valid, high-frequency snapshots of psychobehavioral state throughout the pre-experiment phase (Figure~\ref{fig:long}).

\subsection{In-Lab Data Collection}

\begin{figure}
    \centering
    \includegraphics[width=1\textwidth]{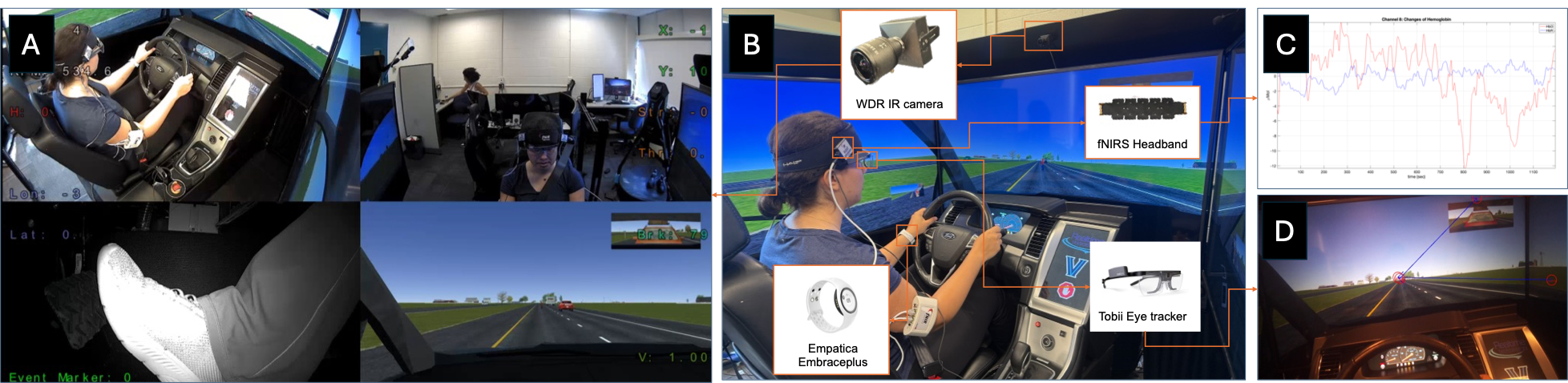}
    \caption{In-lab data collection framework during takeover: (A) video recordings; (B) simulator setup and sensing equipment; (C) fNIRS output sample; (D) gaze output from Tobii glasses showing fixation point}
    \label{fig:in-lab}
\end{figure}

Figure~\ref{fig:in-lab} illustrates the multimodal data collection setup used during the simulator session. The five primary sensing modalities are described below.

\subsubsection{fNIRS: Cortical Activity Monitoring}

\begin{figure}[ht]
    \centering
    \includegraphics[width=0.6\textwidth]{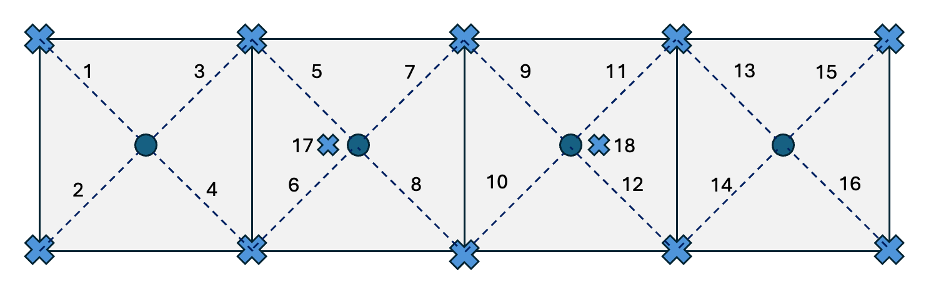}
    \caption{fNIRS source-detector layout over prefrontal cortex}
    \label{fig:FNIRS_layout}
\end{figure}

Cortical activation was measured using the fNIRS Imager 2000 (Biopac Systems, Inc.), which sampled prefrontal cortex (PFC) activity at 2 Hz. The sensor array included four LED light sources (emitting at 730, 805, and 850 nm) and twelve photodetectors, yielding 16 long channels (2.5 cm source-detector separation) and 2 short channels (1 cm). Placement followed standardized protocols, as shown in Figure~\ref{fig:FNIRS_layout}.

\subsubsection{Eye Tracking: Gaze Behavior}

Tobii Pro Glasses 2 recorded binocular gaze at 100 Hz, capturing 2D screen coordinates, 3D gaze intersections, gaze direction vectors, pupil size, and head motion via embedded IMU sensors. Glasses were individually calibrated for accuracy prior to each session.

\subsubsection{Driving Performance Metrics}

The high-fidelity simulator recorded behavioral metrics including speed, lane position, acceleration/deceleration, time headway, and steering wheel angle. These data were used to assess takeover performance and driving quality under different secondary task and takeover context conditions.

\subsubsection{Video Recording: Behavioral Context}

Multiple in-cabin and external cameras captured participants’ upper body posture, hand movements, and facial expressions. A KT\&C Fine Eye 1080p HD camera was positioned at an oblique rear-right angle, while a front-facing WDR camera (IR 2.8–21mm, 3MP) was mounted on the dashboard. These recordings supported behavioral coding and task compliance verification.

\subsection{Data Analysis and Feature Extraction}\label{lab:data_analysis}

\subsubsection{Longitudinal Data}

To characterize baseline physiological state, we analyzed longitudinal data from the Empatica EmbracePlus collected during the 7 days prior to the simulator session. The device’s photoplethysmography (PPG) sensor captures blood volume pulse (BVP), from which inter-beat intervals (IBIs) were extracted to compute heart rate variability (HRV). We focused on the root mean square of successive differences (RMSSD), a time-domain HRV metric linked to parasympathetic activity and associated with emotional regulation, cognitive control, and stress resilience \cite{kim2018stress}. Baseline RMSSD was calculated by averaging values recorded during sleep, identified using Empatica’s automated annotations, to minimize daytime confounds. In parallel, daily self-report data on stress, valence, arousal, and sleep quality were averaged across all days to generate participant-level summaries.

\subsubsection{Functional near-infrared spectroscopy (fNIRS)}
Raw optical intensity data at 730 and 850 nm were first assessed for low amplitude, saturation, and noise contamination using the Scalp Coupling Index (SCI). SCI was computed by band-pass filtering the signal (0.8–2.3 Hz) to isolate cardiac pulsations, followed by calculating the correlation between the two wavelengths \cite{pollonini2016phoebe}. Channels with an SCI below 0.4 were excluded due to poor scalp contact. As anatomical differences (e.g., forehead shape, tissue depth) can affect signal quality, SCI served as an objective criterion for identifying low-quality channels.

After quality assessment, signals were high-pass filtered (Butterworth, cutoff 0.009 Hz) to remove slow drifts. Motion artifacts were corrected using targeted spline interpolation \cite{scholkmann2010detect}, followed by Daubechies wavelet filtering (order 2, type 5) \cite{molavi2012wavelet}. Filtered intensity signals were converted to changes in oxyhemoglobin (HbO) and deoxyhemoglobin (HbR) using the modified Beer–Lambert Law \cite{herff2014mental}, with differential pathlength factors (DPFs) computed by wavelength and age \cite{scholkmann2013general}. Molar extinction coefficients were taken from \cite{omlcTabulatedMolar} as follows: $eHbR_{730} = 1.1022$, $eHbO_{730} = 0.390$, $eHbR_{805} = 0.73708$, $eHbO_{805} = 0.836$, $eHbR_{850} = 0.69132$, and $eHbO_{850} = 1.058$.

\begin{figure}[ht]
    \centering
    \includegraphics[width=0.95\linewidth]{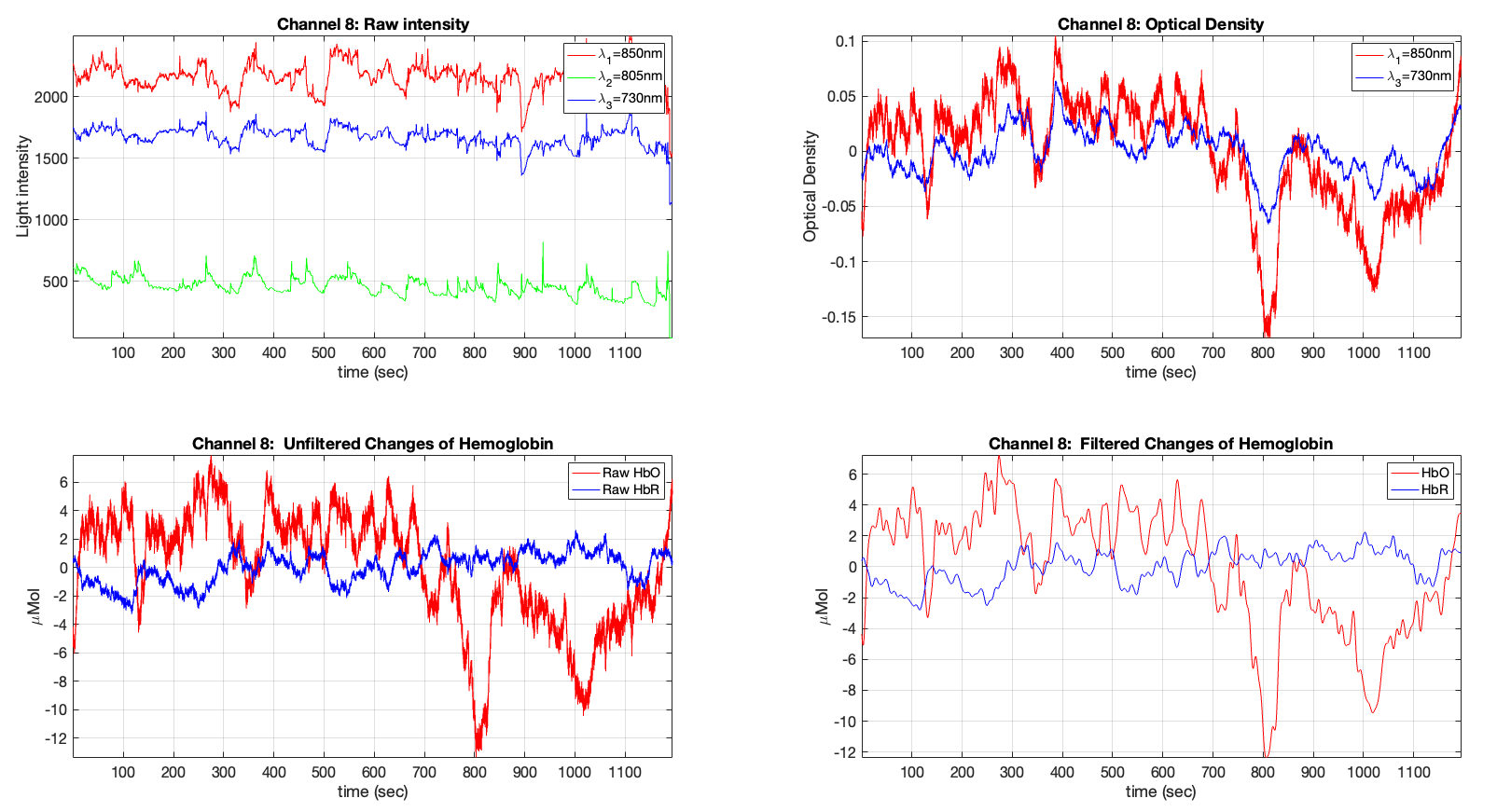}
    \caption{Preprocessing procedure for retrieving clean HbO signal from the fNIR light intensity readings. }
    \label{fig:fnir_pre}
\end{figure}

To reduce physiological noise such as respiration and cardiac pulsation, HbO and HbR signals were low-pass filtered using a Butterworth filter with a 0.10 Hz cutoff. Data epochs were then extracted for each condition—single-task driving (STD) and dual-task driving (DTD)—across all valid long channels (up to 16 per participant, based on SCI). Each epoch was baseline-corrected by subtracting the mean of the preceding 10-second baseline segment. Mean values relative to baseline were computed for HbO and HbR in each condition and channel and used in subsequent analyses. This preprocessing pipeline was applied to each participant’s data, with an example output shown in Fig.~\ref{fig:fnir_pre}.

\subsubsection{Eye Gaze Patterns}

Raw gaze data from the Tobii Pro Glasses 2 were processed using TobiiGlassesPySuite, an open-source Python toolkit for analyzing Tobii outputs \cite{de2019tobiiglassespysuite}. Fixation events were identified using the dispersion-threshold identification (I-DT) algorithm, with fixations defined as clusters of gaze points within 1° of visual angle lasting at least 200 ms, following established conventions \cite{llanes2020development}. The angular threshold was converted into screen-space units to align with the coordinate system of the gaze data. Mean fixation duration was then computed for each participant as a gaze-based behavioral indicator of cognitive demand \cite{broadbent2023cognitive}. Values were calculated separately for three task conditions (No Task, N-Back, and Conversation) and for both takeover scenarios: Unexpected Pedestrian and Crash. Fixation duration was computed as a global measure of gaze behavior across the visual scene and was not segmented by specific areas of interest.

\subsubsection{Driver Takeover Readiness Assessment:}

Driver readiness to retake control was assessed using Takeover Control Time (TOC), a widely used behavioral metric defined as the time between a system-initiated takeover request (TOR) and the driver’s first observable input—such as steering, braking, or accelerating \cite{greer2023safe}. These actions signal re-engagement with the driving task. Our approach aligns with prior work that operationalizes TOC based on physical control inputs \cite{kuehn2017takeover, yang2023takeover}.


\subsubsection{Statistical Modeling}
For preliminary statistical analysis, we used the Wilcoxon signed-rank test for non-parametric pairwise comparisons \cite{woolson2007wilcoxon}, implemented via Python’s scipy.stats package. Holm correction was applied to adjust for multiple comparisons \cite{holm1979simple}. To assess individual variability in longitudinal measures, we used linear mixed-effects models (LMMs) \cite{oberg2007linear}, which are well suited for repeated-measures designs, allowing variance to be partitioned into within- and between-participant components.

\section{Results}


The following section presents a preliminary analysis of data from the longitudinal monitoring phase and simulator-based sessions. We first examine psychophysiological and self-report variables from the pre-experiment period, followed by responses to takeover events and secondary task engagement in the simulator. The aim is to demonstrate the usability and integrative potential of this hybrid framework.

\subsection{Exploration in the Longitudinal Data}

For the longitudinal data, we examined two sets of measures: (1) emotional indicators—stress, valence, and arousal—and (2) sleep-related measures—self-reported sleep quality and RMSSD during sleep.

\begin{figure}[ht]
    \centering
    \includegraphics[width=0.8\linewidth]{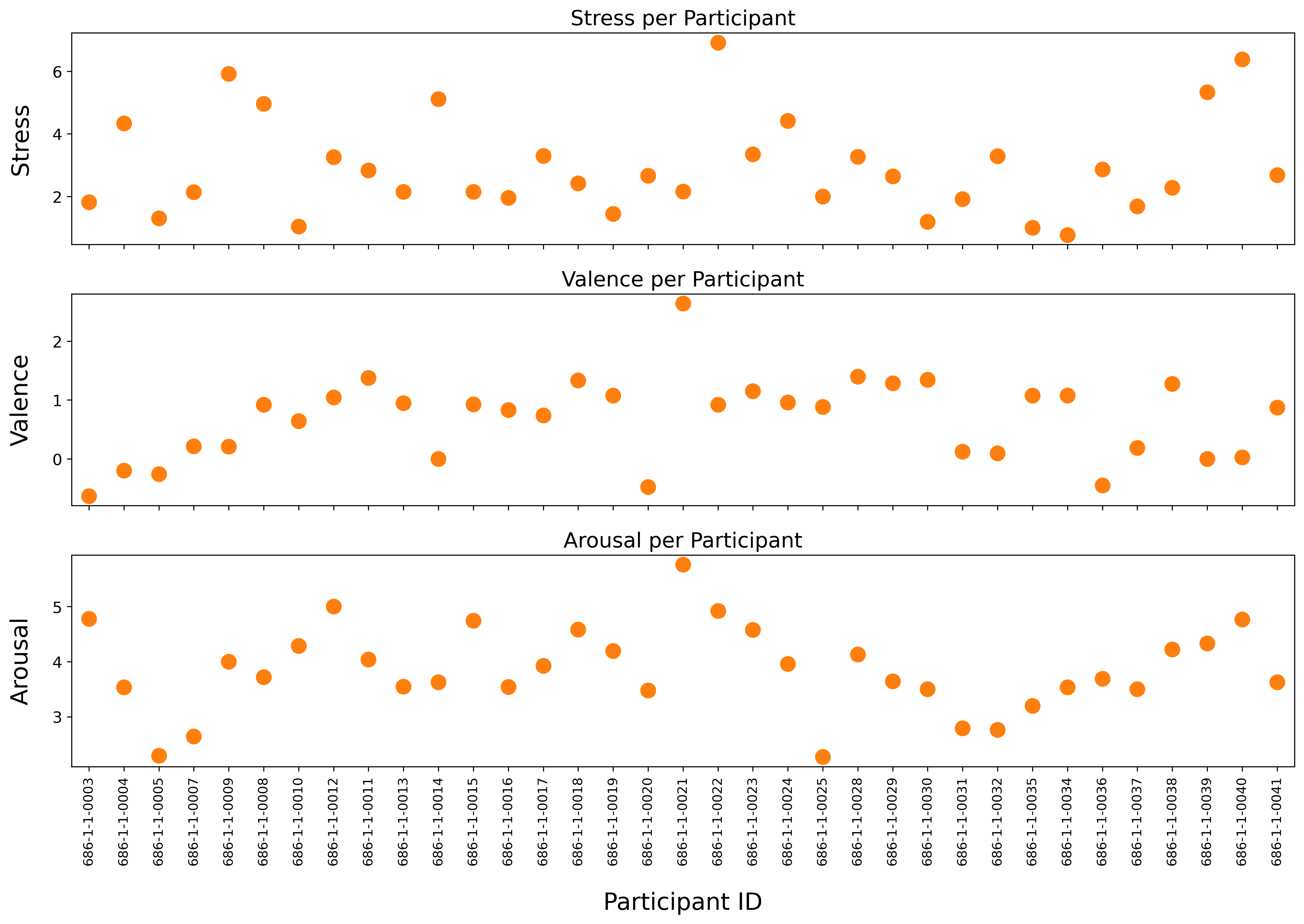}
    \caption{Participants’ average emotional (i.e., stress, valence, and arousal) responses during the study period.}
    \label{fig:emotion-long}
\end{figure}

Figure~\ref{fig:emotion-long} shows average pre-experiment values for stress, valence, and arousal per participant. Valence scores (ranging from –3 to 3) illustrate diverse emotional baselines, with some participants showing consistently positive affect and others exhibiting more neutral or negative patterns.

\begin{figure}[ht]
\centering
\includegraphics[width=0.7\linewidth]{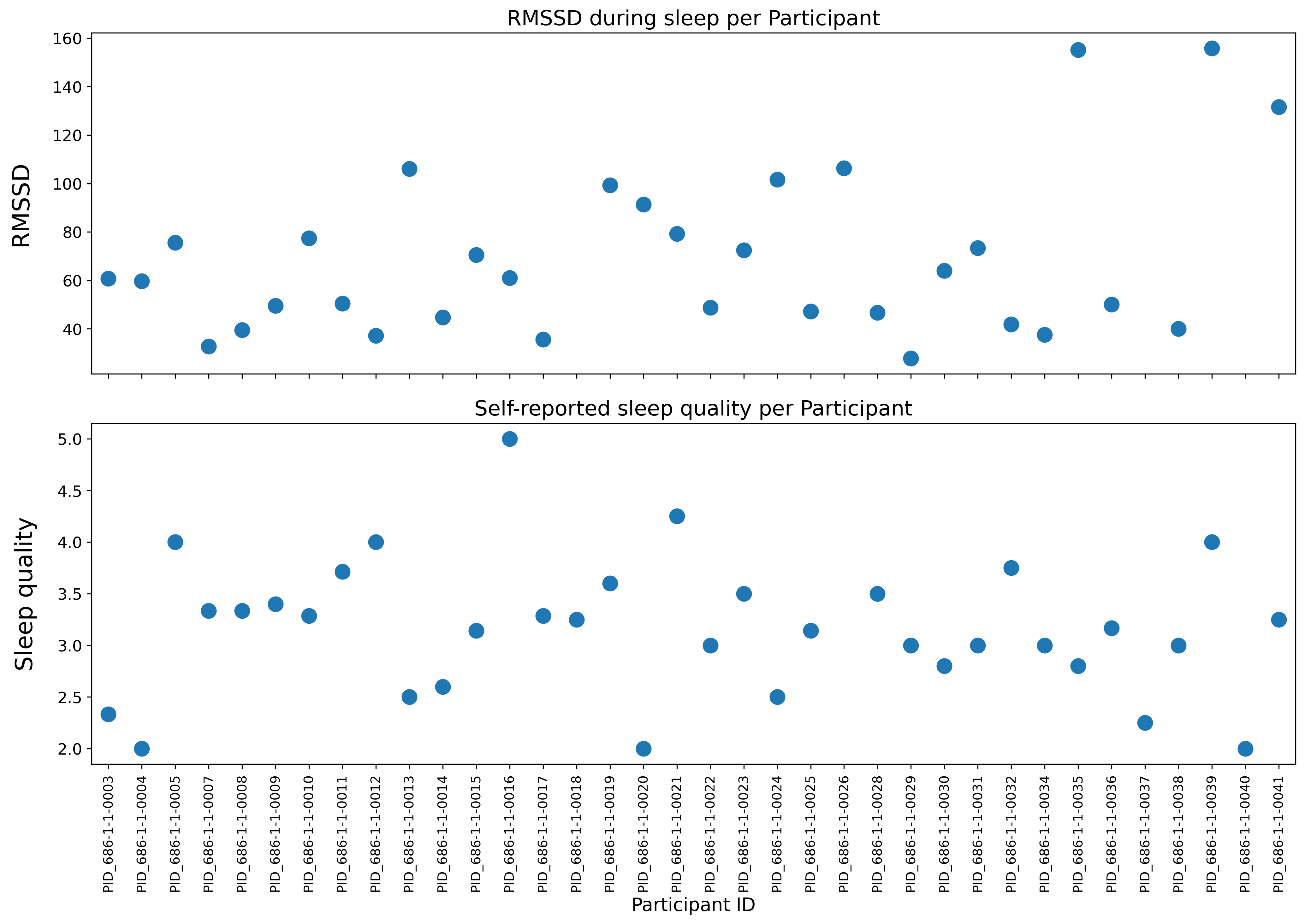}
\caption{Participants’ average RMSSD during sleep (top) and self-reported sleep quality (bottom) across all available participants.}
\label{fig:sleep}
\end{figure}

Figure~\ref{fig:sleep} presents participants’ average RMSSD during sleep (top) and self-reported sleep quality (bottom), illustrating individual variability in both physiological and subjective sleep measures.

To assess individual variability in longitudinal measures, we fit linear mixed-effects models (LMMs) for stress, valence, arousal, sleep quality, and baseline RMSSD. Each model included participant ID as a random intercept to account for repeated observations, enabling variance to be partitioned into between- and within-participant components. LMMs offer a flexible framework for modeling individual differences over time and are well suited for nested, repeated-measures data.

From each model, we computed the Intraclass Correlation Coefficient (ICC) \cite{shrout1979intraclass} to quantify the proportion of variance attributable to between-person differences. ICC was calculated as:
\begin{equation}
\text{ICC} = \frac{\sigma^2_{\text{between}}}{\sigma^2_{\text{between}} + \sigma^2_{\text{residual}}}
\end{equation}

Higher ICC values indicate greater stability across individuals, whereas lower values reflect greater within-person variability over time.

As shown in Table~\ref{tab:lmm_icc_summary}, baseline RMSSD had the highest ICC (0.817), indicating strong between-person consistency in sleep-related physiology. ICCs for psychological self-reports were moderate (stress = 0.388; valence = 0.328; arousal = 0.209), while sleep quality showed the lowest stability (ICC = 0.147), reflecting high day-to-day variability. These results confirm meaningful individual differences and provide a statistical basis for linking baseline states to driving behavior in later analyses.

\begin{table}
\centering
\caption{Summary of Linear Mixed Model Results and ICC Estimates for Longitudinal Measures}
\vspace{6pt}
\resizebox{1\textwidth}{!}{
\begin{tabular}{lcccccccc}
\toprule
\textbf{Measure} & \textbf{N Obs.} & \textbf{N Groups (participants)} & \textbf{Intercept (SE)} & \textbf{z} & \textbf{p-value} & \textbf{Group Var} & \textbf{Residual Var} & \textbf{ICC} \\
\midrule
Stress & 814 & 37 & 2.953 (0.255) & 11.56 & $< 0.001$ & 2.239 & 3.528 & 0.388 \\
Valence & 814 & 37 & 0.742 (0.123) & 6.02 & $< 0.001$ & 0.511 & 1.045 & 0.328 \\
Arousal & 814 & 37 & 3.907 (0.130) & 30.01 & $< 0.001$ & 0.529 & 2.000 & 0.209 \\
Sleep Quality & 157 & 35 & 3.187 (0.100) & 31.99 & $< 0.001$ & 0.144 & 0.838 & 0.147 \\
Baseline RMSSD & 6483 & 37 & 67.165 (5.374) & 12.50 & $< 0.001$ & 1066.670 & 238.211 & 0.817 \\
\bottomrule
\end{tabular}
}
\label{tab:lmm_icc_summary}
\end{table}

\subsection{Exploration in fNIRS Measurements}
\begin{figure}[ht]
\centering
\includegraphics[width=0.8\linewidth]{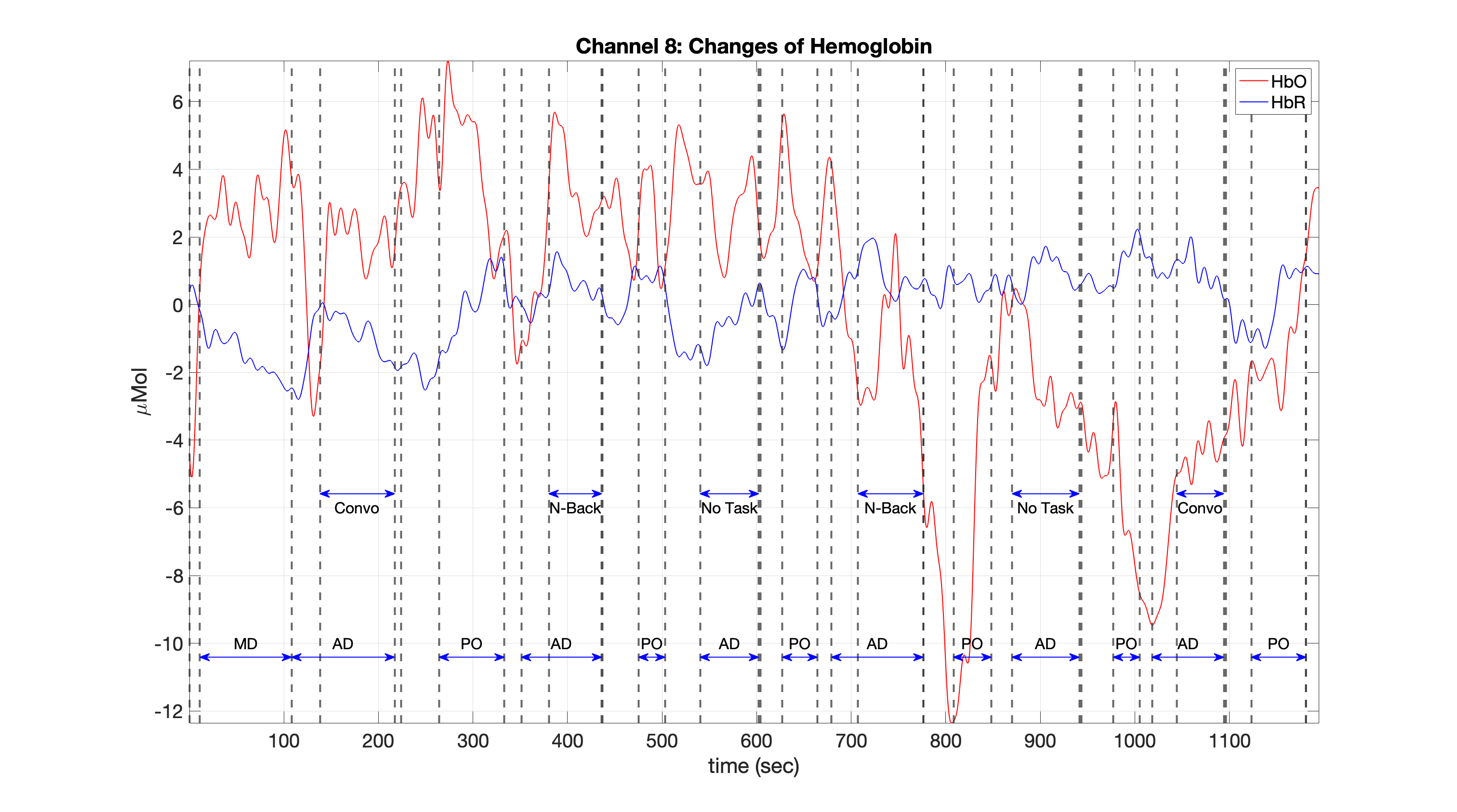}
\caption{Example data from a representative subject showing hemodynamic responses recorded via fNIRS at Channel 8 across different driving and task conditions. Oxyhemoglobin (HbO) is shown in red and deoxyhemoglobin (HbR) in blue. The session included alternating periods of conversational tasks, n-back working memory tasks, and no-task intervals, all performed during autonomous driving (AD).}
\label{fig:fnirs_mkr}
\end{figure}

To illustrate the type of data captured by the fNIRS system, Figure~\ref{fig:fnirs_mkr} shows HbO and HbR signals from Channel 8 for a representative participant, aligned with alternating driving states—manual driving (MD), autonomous driving (AD), and pull-over (PO)—and task conditions, including conversation, n-back, and no-task intervals. Clear HbO increases with corresponding HbR decreases were observed during cognitively demanding periods, particularly during n-back tasks (e.g., 380–440 sec and 700–770 sec), indicating elevated cortical activation. Conversational tasks showed moderate HbO increases, while no-task intervals exhibited more stable or declining patterns. These responses align with expected hemodynamic changes under varying cognitive load.

\subsection{Exploration in the Eye Gaze Patterns Measure}

\begin{figure}[ht]
    \centering
    \includegraphics[width=0.7\linewidth]{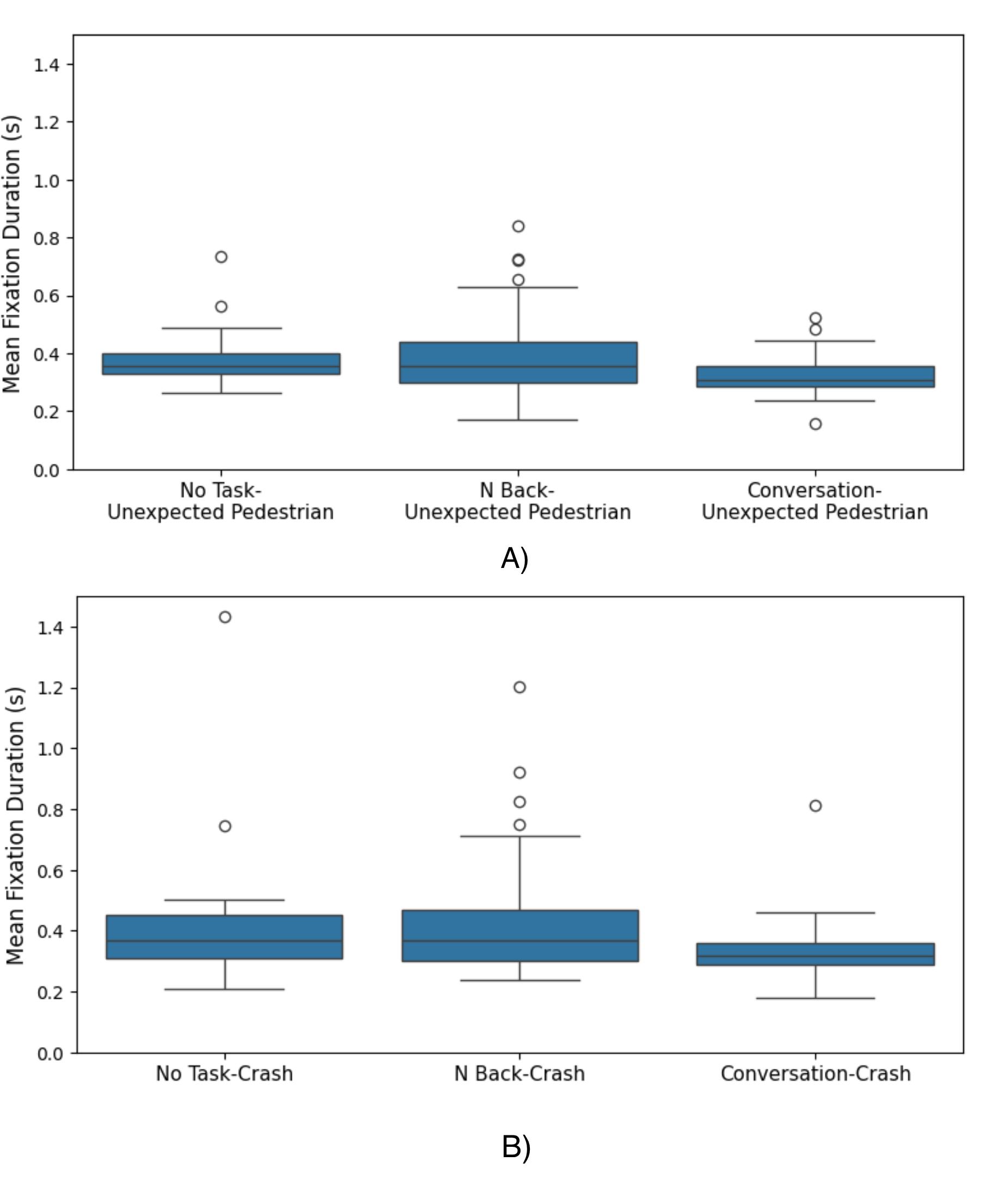}
    \caption{Mean fixation durations across all experimental conditions: No Task, N-Back, and Conversation. A) Unexpected Pedestrian event. B) Crash event.}
    \label{fig:fixation_tasks}
\end{figure}

Figure~\ref{fig:fixation_tasks} shows mean fixation durations across three task conditions—No Task, N-Back, and Conversation—separately for the Unexpected Pedestrian (A) and Crash (B) events. Boxplots indicate task-related variability, with shorter durations observed during the Conversation task. To assess statistical differences, we used Wilcoxon signed-rank tests with Holm correction.

During the Unexpected Pedestrian event, fixation durations in the Conversation task were significantly shorter than in the N-Back (p = 0.0004) and No Task (p = 0.0032) conditions, while no difference was found between N-Back and No Task (p = 0.8879). A similar pattern emerged during Crash events: fixation durations in the Conversation task were significantly shorter than in the N-Back (p = 0.0002) and No Task (p = 0.0065) conditions. These results, summarized in Table~\ref{tab:mean_fixation_wilcoxon_adj_pvalues}, suggest that conversational distraction reduces visual fixation duration, which serves as a gaze-based behavioral indicator of increased cognitive demand.

\begin{table}[ht]
\centering
\caption{Adjusted p-values from Pairwise Wilcoxon Tests (Holm Correction) for Mean Fixation Duration}
\vspace{6pt}
\label{tab:mean_fixation_wilcoxon_adj_pvalues}

\resizebox{\columnwidth}{!}{%
\begin{tabular}{lccc|ccc}
\toprule
& \multicolumn{3}{c|}{\textbf{Unexpected Pedestrian Event}} 
& \multicolumn{3}{c}{\textbf{Crash Event}} \\
\cmidrule(lr){2-4} \cmidrule(lr){5-7}
& \textbf{No Task} 
& \textbf{N-Back Task} 
& \textbf{Conversation Task} 
& \textbf{No Task} 
& \textbf{N-Back Task} 
& \textbf{Conversation Task} \\
\midrule
\textbf{No Task} & -- & 0.8879 & \textbf{0.0032} & -- & \textbf{0.0449} & \textbf{0.0065} \\
\textbf{N-Back Task} & 0.8879 & -- & \textbf{0.0004} & \textbf{0.0449} & -- & \textbf{0.0002} \\
\textbf{Conversation Task} & \textbf{0.0032} & \textbf{0.0004} & -- & \textbf{0.0065} & \textbf{0.0002} & -- \\
\bottomrule
\end{tabular}
}
\end{table}

\subsection{Exploration in the Takeover Control (TOC) Time Measure}

\begin{figure}[ht]
    \centering
    \includegraphics[width=0.8\linewidth]{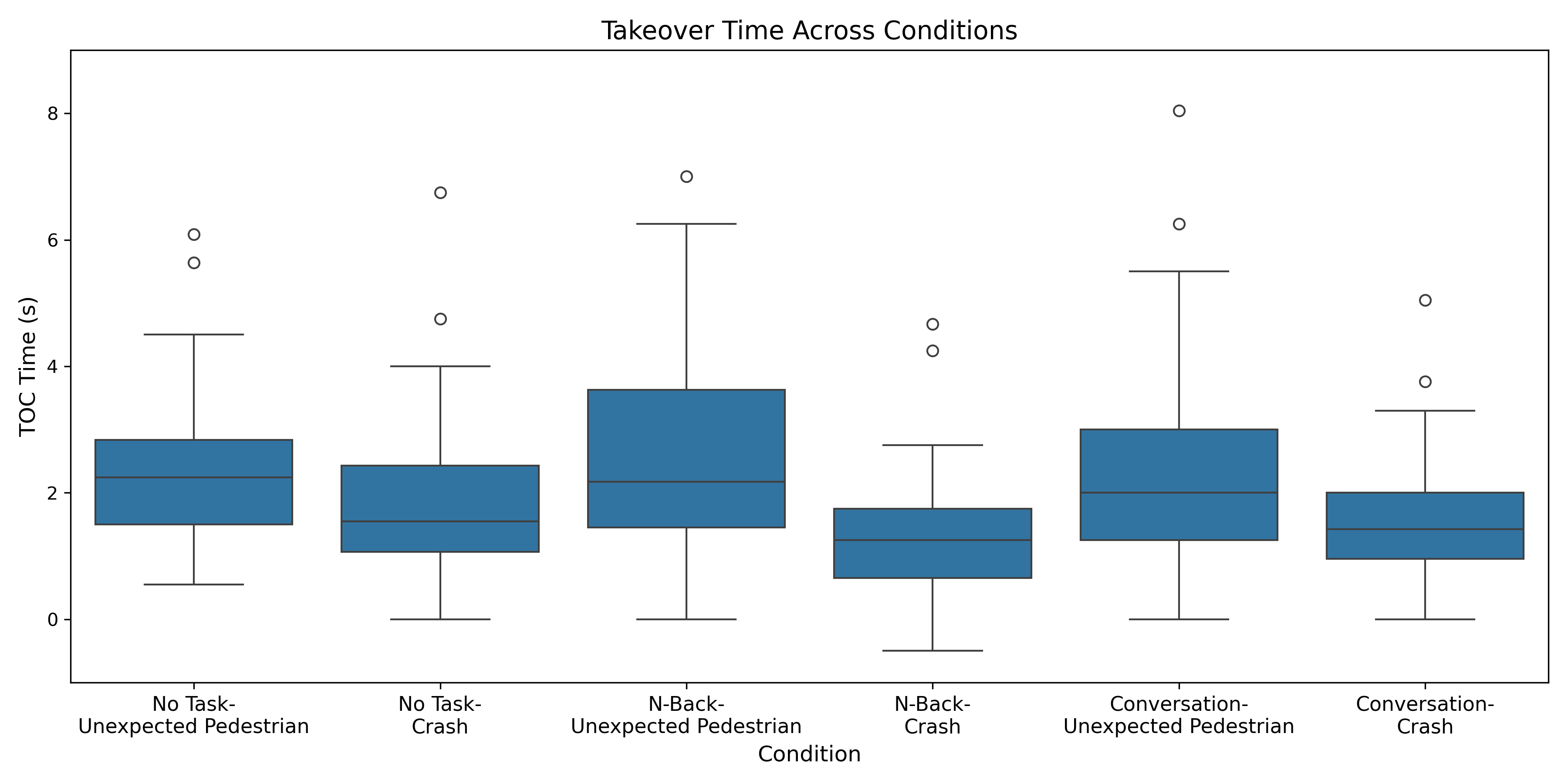}
    \caption{Takeover control (TOC) times across all experimental conditions: No Task, N-Back, and Conversation, each with Unexpected Pedestrian and Crash events.}
    \label{fig:toc_conditions}
\end{figure}

Figure~\ref{fig:toc_conditions} shows takeover control (TOC) times across all conditions, combining No Task, N-Back, and Conversation tasks with either an Unexpected Pedestrian or Crash event. Notably, TOC times were highest in the Conversation–Unexpected Pedestrian condition, highlighting the impact of cognitive distraction on response latency.

To compare TOC times across conditions, we conducted Wilcoxon signed-rank tests with Holm correction, accounting for the within-subjects design and non-normal TOC distributions. Adjusted p-values are shown in Table~\ref{tab:wilcoxon_adj_pvalues}. While several condition pairs were significant, the primary focus was secondary tasks. The \textit{Conversation–Unexpected Pedestrian} condition differed significantly from \textit{Conversation–Crash} (\textit{p} = 0.0057), suggesting that the emotional salience of the event affected performance under conversational distraction. Additionally, the \textit{N-Back–Unexpected Pedestrian} condition differed significantly from both \textit{Conversation–Crash} (\textit{p} = 0.0199) and \textit{No Task–Crash} (\textit{p} = 0.0057).

\begin{table}[ht]
\centering
\caption{Adjusted p-values from Pairwise Wilcoxon Tests (Holm Correction)}
\label{tab:wilcoxon_adj_pvalues}
\resizebox{1\linewidth}{!}{%
\begin{tabular}{lcccccc}
\toprule
 & \textbf{N-Back–Unexpected} & \textbf{N-Back–Crash} & \textbf{Conversation–Unexpected} & \textbf{Conversation–Crash} & \textbf{No Task–Unexpected} & \textbf{No Task–Crash} \\
\midrule
\textbf{N-Back–Unexpected} & -- & 0.0907 & 1.0000 & \textbf{0.0199} & 1.0000 & \textbf{0.0057} \\
\textbf{N-Back–Crash} & 0.0907 & -- & 0.5702 & 0.3915 & 0.5702 & 0.6506 \\
\textbf{Conversation–Unexpected} & 1.0000 & 0.5702 & -- & \textbf{0.0057} & 1.0000 & \textbf{0.0153} \\
\textbf{Conversation–Crash} & \textbf{0.0199} & 0.3915 & \textbf{0.0057} & -- & 0.3915 & 0.6506 \\
\textbf{No Task–Unexpected} & 1.0000 & 0.5702 & 1.0000 & 0.3915 & -- & 0.0669 \\
\textbf{No Task–Crash} & \textbf{0.0057} & 0.6506 & \textbf{0.0153} & 0.6506 & 0.0669 & -- \\
\bottomrule
\end{tabular}%
}
\end{table}


\section{Discussion}

This study presents a hybrid framework that integrates longitudinal mobile sensing with high-fidelity driving simulation to examine behavior in semi-automated vehicle contexts. Rather than focusing on prediction, our goal was to assess the feasibility of capturing individual variability across psychophysiological, behavioral, and neural dimensions. The integration of baseline measures (e.g., sleep, stress, RMSSD) with controlled in-lab data enables observation of both real-time responses and pre-existing states. This approach lays the groundwork for future research on how temporal dynamics and individual differences shape driver readiness.

The longitudinal component of this framework was particularly useful for capturing both day-to-day fluctuations and stable patterns in psychophysiological states. Despite some missing data, multi-day measures of stress, valence, arousal, sleep quality, and RMSSD were successfully collected for most participants. Linear mixed-effects models revealed moderate to high ICCs, with RMSSD during sleep showing the highest stability, suggesting its value as a trait-level indicator. While direct links to driving performance were not tested here, these findings demonstrate the feasibility of integrating mobile sensing into driver research and underscore the potential for personalizing driver state models based on baseline variability. This finding is consistent with prior driving and human-factors research identifying RMSSD as a robust indicator of autonomic regulation and cognitive workload in simulated and automated driving contexts, and one that is well suited for characterizing stable individual differences \cite{mehler2009physiological,stapel2019automated}.

We incorporated neural sensing via fNIRS during the simulator session. Although only one participant's data is presented in this preliminary analysis, the results indicate fNIRS can detect task-related fluctuations in brain activity. Increases in oxygenated hemoglobin were observed during cognitively demanding conditions, particularly the n-back task, and to a lesser extent during conversation—consistent with expected prefrontal activation patterns \cite{yeung2023changes}. This adds a valuable neural dimension to the framework and supports future work linking brain activity to takeover performance.

Gaze behavior showed that task demands influenced visual attention. Fixation durations were shorter during the Conversation condition compared to No Task and N-Back, suggesting more fragmented visual attention under conversational distraction \cite{ries2016impact}. In contrast, the N-Back task was associated with longer fixation durations, consistent with increased working memory demands \cite{keskin2020exploring}. These findings align with prior research indicating that fixation duration serves as a gaze-based behavioral indicator of cognitive demand and varies with task difficulty \cite{bitkina2021}.

TOC time analysis shows this framework can assess driver responses under varying task-induced cognitive and contextual demands. TOC times differed significantly across conditions, with the Conversation–Unexpected Pedestrian scenario producing the longest delays. This aligns with prior research showing that naturalistic conversation impairs situational awareness and delays response initiation \cite{rakauskas2004effects}. This mirrors gaze findings, where conversational tasks reduced fixation durations, indicating conversational distraction degrades both perceptual and behavioral readiness.
Prior driving research suggests that the disruptive effects of conversational secondary tasks are amplified in dynamic and unexpected driving events, where increased event salience and time pressure elevate attentional reorientation demands under dual-task conditions \cite{strayer2007cell,wilson2008driver}; however, because takeover contexts were presented in a fixed order, these effects should be interpreted cautiously and examined further using counterbalanced designs.

TOC variability was shaped not only by the type of secondary task but also by the nature of the takeover event, suggesting an interaction between task-induced cognitive load and external context. Conversation generally delayed responses, especially in the emotionally salient Unexpected Pedestrian condition, whereas task-related differences were less pronounced during crash events. This implies that perceived urgency or surprise may modulate the impact of distraction. However, since event type was not randomized within participants, these effects may also reflect scenario order or expectations. While this limits interpretive strength, the findings raise important questions about how internal and external factors jointly influence takeover readiness. Future studies with counterbalanced designs are needed to test these interactions more rigorously. Still, the results support the value of multimodal sensing to capture the layered complexity of driver behavior in semi-automated systems.

\section{Limitations and Future Work}

This study, while demonstrating the feasibility of a hybrid framework, faced several logistical and methodological challenges. In the longitudinal phase, participant adherence to wearable use—especially during sleep—was inconsistent, resulting in missing physiological data such as nighttime RMSSD. Daily survey compliance also varied across participants, despite reminders and streamlined interfaces. These issues highlight the need for more effective engagement strategies in future multi-day studies. In the simulator phase, aligning data streams from fNIRS, eye tracking, and physiological sensors was challenged by minor hardware and software delays. Improved real-time sensor integration protocols will be important for scaling this approach.

Additionally, the gaze analysis used only fixation duration, which does not capture the full complexity of visual attention. Future studies should incorporate additional metrics such as saccade dynamics, blink rate, and pupil dilation, and expand the analysis window to include transitions between driving modes. The study’s relatively small and demographically homogenous sample—primarily university students—also limits generalizability. This study was intentionally designed as a pilot, proof-of-concept investigation to assess the feasibility of the proposed multimodal framework, validate experimental protocols, assess data integration across modalities, and characterize patterns of variability prior to large-scale empirical modeling. Broader recruitment across age groups and driving backgrounds is critical for developing inclusive driver models. Finally, the use of scripted takeover events limited the ecological richness of the scenarios. Building on this framework, future work will extend the investigation of takeover behavior to more varied and naturalistic scenarios, including Virtual Reality (VR) environments, to support scalable data collection and improve real-world applicability.

\section{Conclusion}
This study introduced a hybrid framework combining longitudinal mobile sensing with high-fidelity driving simulation to examine human readiness in semi-automated vehicles. By capturing both baseline psychophysiological states and real-time responses to takeover events, the framework offers a personalized, temporally sensitive view of driver behavior. The results demonstrate feasibility, reveal individual variability, and underscore the importance of adaptive systems that can anticipate and support human drivers as automation becomes more prevalent.

\section*{Acknowledgment}

The authors would like to thank the Villanova Institute for Research and Scholarship (VIRS) for supporting this research through the Research Catalyst Grant (RCG). The authors also thank the College of Engineering, Center for Human-Environmental Systems (CHES), and the Department of Civil and Environmental Engineering at Villanova University for supporting the driving simulator used in our study.




\bibliographystyle{IEEEtran}
\bibliography{IEEEexample}

\end{document}